\documentclass[aps,twocolumn]{revtex4}
\usepackage{epsfig}

\newcommand{\be}{\begin{equation}}
\newcommand{\ee}{\end{equation}}
\newcommand{\ba}{\begin{array}}
\newcommand{\ea}{\end{array}}

\begin{document}

\title{Collapsed 2-Dimensional Polymers on a Cylinder}

\author{Hsiao-Ping Hsu and Peter Grassberger}

\affiliation{John-von-Neumann Institute for Computing, Forschungszentrum J\"ulich,
D-52425 J\"ulich, Germany}

\date{\today}

\begin{abstract}
Single partially confined collapsed polymers are studied in two
dimensions. They are described by self-avoiding random walks
with nearest-neighbour attractions below the $\Theta$-point, on
the surface of an infinitely long cylinder. For the simulations
we employ the pruned-enriched-Rosenbluth method (PERM).
The same model had previously been studied for free
polymers (infinite lattice, no boundaries) and for
polymers on finite lattices with periodic boundary conditions.
We verify the previous estimates of bulk densities, bulk
free energies, and surface tensions. We find that the
free energy of a polymer with fixed length $N$ has, for $N\to
\infty$, a minimum at a finite cylinder radius $R^*$ which
diverges as $T\to T_\theta$. Furthermore, the surface tension 
vanishes roughly as $(T_\theta-T)^\alpha$ for $T\to T_\theta$
with $\alpha\approx 1.7$. The density in the interior of a globule
scales as $(T_\theta-T)^\beta$ with $\beta \approx 0.32$.
\end{abstract}

\maketitle

Although the behaviour of polymers has been studied for many years,
there are still a number of open questions.
At high temperatures or in good solvents repulsive interactions
(the excluded volume effect) dominate the conformation
and the polymer chain tends to swell to
a random coil. At low temperatures or in poor solvents attractive
interactions between monomers dominate the conformation and
the polymer chain tends to collapse and forms a compact dense globule.
The coil-globule transition point is called the $\Theta$ point.

The behaviour in good solvents is accessible by field theory and thus
well understood \cite{deGen}, e.g. the partition function scales
as $Z \sim \mu^N N^{\gamma-1}$ at $T>T_\theta$. In poor solvents,
much less is known exactly. For a long time it was e.g. widely believed
that the partition sum scales in the same way, although already
Lifshitz mean-field theory \cite{Grosberg,Lifshitz} had predicted
for $T<T_\theta$ that there should be also a surface term,
\be
     Z(N,T) \sim a^N b^{-N^s} N^{\gamma-1}\;,  \label{Z}
\ee
with $s=(d-1)/d$ and $b>1$. First numerical indications for this 
were given by \cite{Owczarek,hegger,nidras}, and theoretical arguments
in favour of Eq.~(\ref{Z}) were given in \cite{duplantier}. More recently,
Eq.~(\ref{Z}) was checked numerically in \cite{g02}. But it seems that
there exists still no proof for Eq.~(\ref{Z}), and it is not known
how the surface tension (related to $b$) scales for $T\to T_\theta$.
Moreover, the exponent $\gamma$ is not known \cite{Owczarek,duplantier,nidras},
and it is not even clear whether it is universal in the collapsed phase.

In \cite{g02} we had studied two different geometries -- globules on
infinite lattices and polymers filling finite lattices with periodic
boundary conditions -- where surface tension should make very different
contributions. In the present work we study a third geometry, namely
polymers on the surface of an infinitely long cylinder, where the surface
tension should make yet another contribution. Our main motivation was
originally to have a simple geometry with straight ``surfaces" so that
the surface tension and its $T$-dependence can be more easily measured.
But we found also another interesting effect: the bulk fugacity $\mu$
depends non-monotonically on the cylinder radius. It has a minimum at a
finite radius $R^*$ which diverges for $T\to T_\theta$. Thus a collapsed
polymer, when wrapped around an elastic cylinder, will tend to squeeze
it when $R>R^*$ and widen it when $R<R^*$.

\begin{figure}
  \begin{center}
\psfig{file=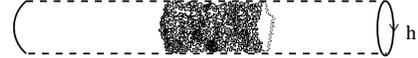,width=5.4cm, angle=0}
   \caption{Schematic drawing of a polymer chain growing on the surface
of a cylinder.}
    \label{cylinder}
  \end{center}
\end{figure}

    A collapsed polymer on the surface of an infinitely long cylinder
is modeled as an interacting self-avoiding random walk (ISAW) on a
square lattice with periodic boundary conditions in the vertical
direction, as shown in Fig.~\ref{cylinder}. The vertical lattice size
is $h=2\pi R$. The collapse is mediated by nearest
neighbour attraction between monomers, described by a Boltzmann
factor $q=e^{-\beta \epsilon}$ for each pair of non-bonded monomers
occupying nearest neighbour sites on the lattice. Here $\beta=1/kT$
and $\epsilon < 0$ is the attractive potential. The partition sum is
therefore
\be
   Z=\sum_{walks}q^{m}\; ,                     \label{SAW}
\ee
where $m$ is the number of non-bonded occupied nearest-neighbour pairs.
We are interested in polymers which form collapsed globules, therefore
$q > q_\theta$ with $q_\theta = e^{-\epsilon/kT_\theta} \approx 1.95$
\cite{barkema}.

As argued in \cite{Grosberg,Lifshitz} and verified in \cite{Owczarek,g02},
below the $\Theta$ point, the polymer forms a globule with
uniform monomer density $\rho$ for large $N$ and $\rho \rightarrow 0$ as
$T \to T_\theta$.
Thus, the free energy for a collapsed polymer with chain length
$N$ in infinite volume consists (except for a logarithmic term resulting
from the last factor in Eq.~(\ref{Z})) of two parts: An extensive bulk
contribution $\propto N$, and a surface contribution $\propto N^{(d-1)/d}$.
For $d=2$, this reads (up to a factor $1/T$ which will be suppressed in the
following)
\be
   - \ln Z_N(q) \approx \mu_\infty(q) N + \tilde{\sigma}(q) \sqrt{N}
              +O(\ln N)     \label{Z2N}
\ee
where $\mu_\infty$ is the chemical potential per monomer in an infinite chain,
$\tilde{\sigma}$ is related to the surface tension (=free energy per
unit of perimeter length) $\sigma$ and to the monomer density $\rho$ by
\be
    \sigma=\frac{1}{2} \tilde{\sigma} \sqrt{\frac{\rho}{\pi}} \; .
\ee
The latter follows by assuming the globule to be a disk with radius $R_g$
fixed by $N=\pi R_g^2\rho$.

Surface terms are absent from the free energy if one has a finite $L\times L$
lattice with periodic boundary conditions, and if one chooses the polymer length
such that the entire lattice is filled with finite density. More precisely,
consider the quantity $ \ln Z_N(q) + \mu N$. For
$T< T_\theta$ this is non-convex, and for $\mu = \mu_L(q)$ it has two peaks
of equal height, corresponding to the coexistence of two phases: One phase
with very short chains and density $\approx 0$, and another phase with density
$\rho_L(q)$. For $L\to \infty$, $\mu_L(q)$ and $\rho_L(q)$ tend
monotonically \cite{g02} to $\mu_\infty(q)$ and $\rho$.

Finally, on the surface of an infinitely long cylinder with perimeter $h$, 
a collapsed polymer forms a cylindrical blob with length $L = \rho^{-1} N
/h$ and surface free energy $2\sigma h$,
\be
  - \ln Z_N(q,h) \approx \mu_h(q) N + 2\sigma(q) h \; .
                   \label{Zh}
\ee
Notice that the chemical potential per monomer will in general depend
on $h$, even in the limit $N\to\infty$. For $h\to\infty$ we expect again to
obtain the bulk limit,
\be
   \mu_h(q) \to \mu_\infty(q) \qquad h\to\infty\;.
\ee
In principle we should also expect $\sigma(q)$ and $\rho(q)$ to depend
on $h$. But we shall not study the detailed behaviour of $\rho$, and we
shall neglect the $h$-dependence of $\sigma(q)$, since the surface term
is a small correction anyhow.
We also might expect logarithmic terms in $N$ to be present (just as
also in the previous geometries), but they will also be neglected.

Our aim is to verify Eq.~(\ref{Zh}) numerically, to estimate the
functions $\mu_h(q)$ and $\sigma(q)$, and to see whether they are
compatible with the results for the other two geometries.

The pruned-enriched-Rosenbluth method (PERM)
has been employed to study collapsed 2-d polymers in \cite{barkema,g02}.
In \cite{g02} the predicted behaviour was verified numerically for globules
on infinite lattices and for polymers filling finite lattices with periodic
boundary conditions. More precisely, finite lattices were studied first,
and from them estimates were obtained for $\mu_\infty$ and $\rho$ by
extrapolation. Given these, subsequent simulations on infinite lattices
gave rather precise estimates for $\sigma(q)$. But the latter was not done
systematically for a wide range of temperatures, in particular the behaviour
of $\sigma(q)$ for $q\to q_\theta $ was not studied.
Thus we want to obtain also the latter in the present paper from cylinder
geometry simulations.

\begin{figure}
  \begin{center}
   \psfig{file=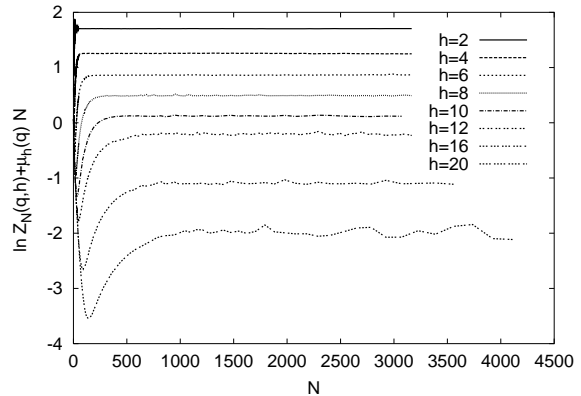,width=5.4cm,
angle=270}
   \caption{$\ln Z_N(q,h)+\mu_h(q)N$ versus $N$ for $q=2.4$ and for various
    values of the cylinder circumference $h$. The values of $\mu$ were
    determined by demanding that these curves become horizontal as $N \to \infty$.
    The peaks seen in the curves for $h=16$ and 20 are statistical fluctuations.}
   \label{lnz}
  \end{center}
\end{figure}

For the first simulations, we choose $q=2.4$ which is deep inside the
collapsed regime.  From Eq.~(\ref{Zh}) we expect that plots of
$\ln Z_N(q,h)+\mu N$ against $N$ should give horizontal lines
for $N \rightarrow \infty$, if and only if $\mu = \mu_h(q)$. This gives
estimates of $\mu_h(q)$. Using them, we plot in Fig.~\ref{lnz}
$\ln Z_N(q,h)+\mu_h(q) N$ versus $N$ for various values of the
cylinder circumference $h$. For small $h$ we find perfect agreement with
our expectations. For larger $h$ we see that PERM has problems in
sampling long chains correctly, as should be expected. Fluctuations similar
to those seen in the curve for $h=16$ prevent us from going to much larger
$h$, except for $q$ close to $q_\theta$ where the efficiency of PERM
increases.

\begin{figure}
  \begin{center}
   \psfig{file=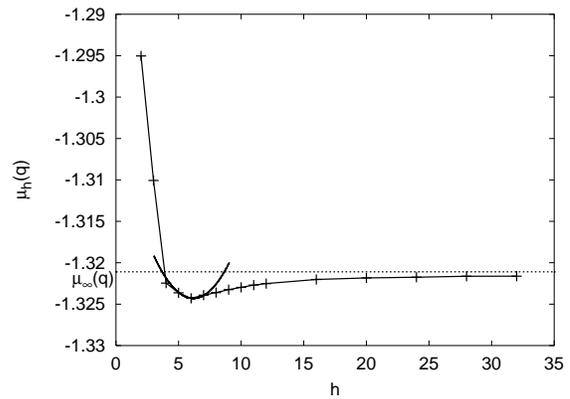,width=5.4cm,angle=
270}
   \caption{$\mu_h(q=2.4)$ as obtained from Fig.~\ref{lnz}, plotted against $h$.
    There is a minimum at $h=6.13 \pm 0.01$ which is determined by fitting
    a parabola (dashed curve) going through the three lowest points (which 
    are at $h=5$, $6$, and $7$). The dotted line is the asymptote  
    $\mu_\infty(q)=-1.3211(1)$ approached for $h\to\infty$.}
   \label{q240mu}
  \end{center}
\end{figure}

Values of $\mu_h(q)$, again for $q=2.4$, are plotted in Fig.~\ref{q240mu}.
For large $h$, we obtain $\mu_\infty(q) = -1.3211(1)$ in perfect
agreement with \cite{g02}. But, surprisingly, the dependence of $\mu_h(q)$ on
$h$ is not monotonic. Instead, we see a minimum at a finite value of $h$.
In our simulations $h$ is an integer, but we can interpolate to
continuous values by fitting a parabola, $\mu_h(q)=\mu_{\rm min}+b(h-h^*)^2$,
through the three lowest points, as shown in Fig.~\ref{q240mu}. Such a
fit gives $h^*(q=2.4)=6.13(1)$ and $\mu_{\rm min}(q=2.4)=-1.3243(1)$.

\begin{figure}
  \begin{center}
   \psfig{file=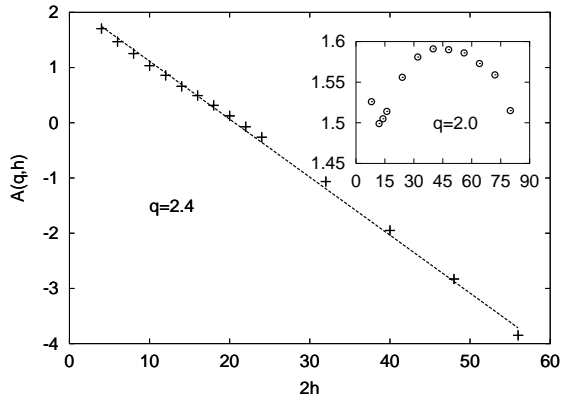,width=5.4cm,
angle=270}
   \caption{$A(q,h)$ versus $2h$
for $q=2.0$ and $2.4$. The dashed line has slope -$\sigma=-0.105$.
The points for $h=16$ and $h=20$ have large errors for $q=2.4$, 
as also seen from Fig.~\ref{lnz}.
Results for $q=2.0$ (insert) show the
difficulty of measuring surface tension near $\Theta$-point.}  
    \label{q240sigma}
   \end{center}
\end{figure}

Finally, in order to obtain the surface tension, we plot in Fig.~\ref{q240sigma}
\be
   A(q,h) = \lim_{N \rightarrow \infty}  [\ln Z_N(q,h)+\mu_h(q) N]
\ee
against $2h$. The limit is of course obtained by using the
parts of the curves in Fig.~\ref{lnz} where they are horizontal.
From Eq.~(\ref{Zh}) we expect this to give a straight line
with slope $-\sigma$. The result shows that $\sigma=0.105(5)$ at $q=2.40$.
Actually, the line is slightly curved, showing that systematic errors are
not negligible. For large $h$ they seem to arise from insufficient sampling.
For small $h$ they result from the neglected terms discussed above.
The present estimate should be compared to the estimate $\sigma=0.119(5)$
from \cite{g02}. Although the discrepancy is about 2 standard
deviations, we consider this as reasonable agreement. Results for $q=2.0$
are also shown in Fig.~\ref{q240sigma}, and illustrate the
large corrections near the $\Theta$-point. 

\begin{figure}
  \begin{center}
   \psfig{file=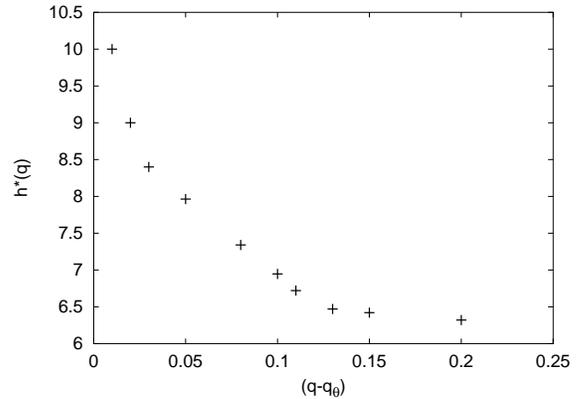,width=5.4cm, angle=270}
   \caption{``Optimal" cylinder circumference $h^*(q)$ against $(q-q_\theta)$.
   At $h=h^*(q)$, the chemical potential is minimal, $\mu_{h^*}(q)=\mu_{\min}(q)$.}
   \label{hmin}
   \end{center}
\end{figure}

\begin{figure}
  \begin{center}
 \psfig{file=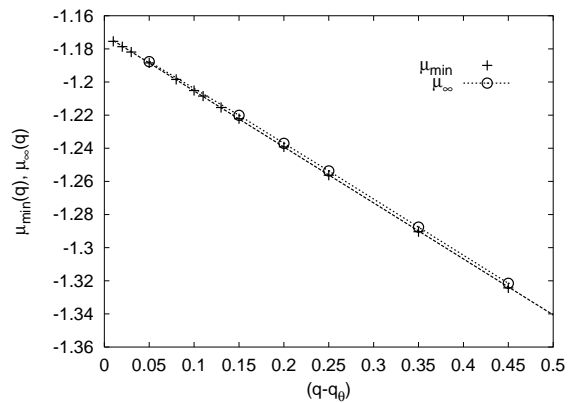,width=5.4cm, angle=270}
   \caption{$\mu_{\rm min}$ and $\mu_\infty$ plotted
against $(q-q_\theta)$ for various values of $q$. The slope of the
straight line is roughly equal $-0.34$.
$\mu_\infty$ were obtained from extrapolation to large $h$.}
   \label{mumin}
  \end{center}
\end{figure}

\begin{figure}
  \begin{center}
  \psfig{file=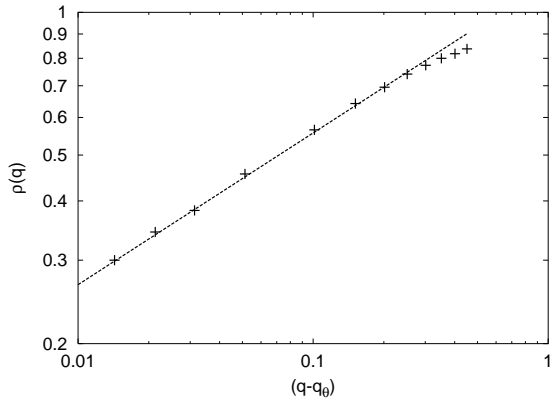,width=5.4cm, angle=270}
   \caption{Log-log plot of $\rho$, as obtained from the growth of 
   $\langle |x_N|\rangle$ with $N$ for large $N$ and $h$, against
   $(q-q_\theta)$. For this plot we used $q_\theta=1.9487$. The dashed line 
   is $\rho \propto (q-q_\theta)^{0.32}$.}
   \label{density}
  \end{center}
\end{figure}

\begin{figure}
  \begin{center}
  \psfig{file=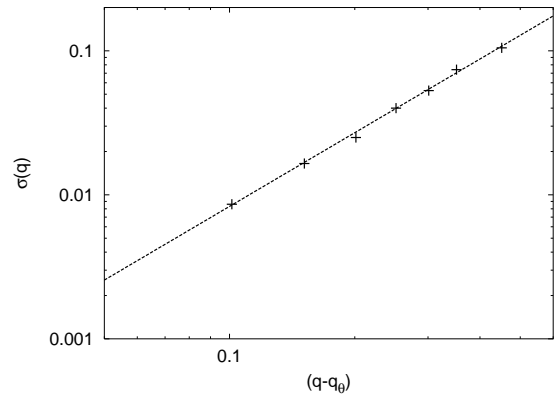,width=5.4cm, angle=270}
   \caption{Log-log plot of $\sigma$, as obtained from Fig.~\ref{q240sigma}, against
   $(q-q_\theta)$. For this plot we used $q_\theta=1.9487$. The dashed line 
   is $\sigma \propto (q-q_\theta)^{1.7}$.}
   \label{sigma}
  \end{center}
\end{figure}

Analogous simulations as shown in Fig.~\ref{lnz}-\ref{q240sigma} were also
done at other values of $q$. The lengths $N$ depended strongly on $q$.
Close to the $\Theta$-point the algorithm is very efficient and we obtained
reliable high statistics data for $N$ up to 15,000. The more the polymer is 
collapsed, the more difficult are the simulations. We stopped at $q=2.4$. 
As seen from Fig.~2, the errors decrease when $h$ becomes small, most likely 
because the walk is effectively Markovian with memory $\sim h^2$, and PERM 
is most efficient when memory is short. In all those simulations the results
were similar. $\mu_h(q)$ had a minimum at a finite $h^*(q)$, and
$A(q,h)$ versus $h$ was roughly linear with a negative slope.
Results of $h^*(q)$, $\mu_{\rm min}(q)$, $\rho(q)$ and $\sigma(q)$ for a wide
range of $q$ are shown in Fig.~\ref{hmin}-\ref{sigma}.

In Fig.~\ref{hmin} we can see that $h^*(q)$ increases as $q$ decreases. 
This is to be expected since the $\Theta$ transition is continuous, with
a diverging correlation length for $T\to T_\theta$. Apart from the lattice 
constant (which is set to 1 here), the correlation length is the only 
relevant length scale. 
The divergence of $h^*(q)$ when approaching the $\Theta$ point thus 
indicates that the existence of the
minimum is not a lattice artifact and is presumably a genuine feature of
the continuum theory. The simulations are not precise enough to allow a
meaningful fit to a power law and a determination of a critical
exponent. This exponent would also be effected strongly by the uncertainty
of $T_\theta$. To reduce the latter to a minimum, we performed extensive 
simulations of free (i.e. infinite lattice) ISAWs near the estimated 
$\Theta$ point, with the result $q_\theta=1.9487(5)$ which is in good
agreement with the best previous estimate \cite{barkema}.

In Fig.~\ref{mumin} we plot $\mu_{\rm min}(q)$ and $\mu_\infty(q)$ versus
$(q-q_\theta)$. We see that both decrease linearly with temperature.
The difference between them depends weakly on the temperature. It seems
to vanish for $T\to T_\theta$.

The monomer density $\rho$ inside the ``globule" is constant for $N \gg h^2 \gg 1$ 
and fixed temperature. It can be measured by measuring the $x$-coordinate
(parallel to the cylinder) of the end point of the chain. Since the 
chain has to grow either into the positive or into the negative direction,
we have 
\be
   {d \langle |x_N|\rangle \over dN } \sim (h \rho )^{-1},\qquad N\to\infty.
\ee
For small values of $h$ there are finite size corrections to the density,
but it is easy to extrapolate to $h\to\infty$. The resulting densities are 
plotted versus $q$ in Fig.~\ref{density}. They show a power law
\be
   \rho \sim (q-q_\theta)^\beta
\ee
with $\beta = 0.32\pm 0.02$. We are not aware of previous estimates of 
$\beta$, either theoretically or by simulations.

The values of the surface tension $\sigma(q)$ obtained by the same 
way as shown in Fig.~\ref{q240sigma} are shown in Fig.~\ref{sigma}. 
They follow a power law, $\sigma \sim (q-q_\theta)^\alpha$, with 
$\alpha\approx 1.7\pm 0.1$. For the parameter $\tilde{\sigma}$ this implies
that
\be
   \tilde{\sigma} \sim (q-q_\theta)^{\alpha-\beta/2} \sim (q-q_\theta)^{1.54\pm 0.1}.
\ee
This is incompatible with the theoretical prediction 
$\tilde{\sigma} \sim (T_\theta-T)^{7/6}$ quoted in \cite{nidras}. 

In summary, we have studied collapsed polymers on the surface of a
cylinder. Using the PERM algorithm we could simulate chains of lengths
between $3000$ (far below $T_\theta$) and $10^4$ (near $T_\theta$) 
for a wide range of temperatures.
We have verified that the concept of a surface tension
applies to 2D collapsed polymers, and have determined its scaling near
the $\Theta$ point. We also determined the scaling of the monomer
density in the collapsed phase, as $T\rightarrow T_\theta$.
In addition, we found a surprising non-monotonic dependence of the
chemical potential on the perimeter of the cylinder. It shows that
for any temperature there is a special value of the perimeter where the
free energy of the polymer is minimal. The fact that this value
diverges when the $\Theta$-temperature is approached suggests that this
is not an artifact of the square lattice but a generic feature.

Acknowledgements: We thank Dr.~Walter Nadler for valuable discussions.
Parts of the simulations were done on a PC-farm in Taipei.
H.~P.~thanks the Computing Centre of Academia Sinica in Taipei, Taiwan
for providing the computing facilities.

\end{document}